\documentclass[aps,prb,showpacs,twocolumn,superscriptaddress,showkeys]{revtex4-1}
\usepackage{amssymb}
\usepackage[dvips]{graphicx}
\usepackage{dsfont}

\begin{document}

\title{Disordered graphene and boron nitride in a microwave tight-binding analog}

\author{S.~Barkhofen}
\affiliation{Fachbereich Physik der Philipps-Universit\"{a}t Marburg, D-35032 Marburg, Germany}
\author{M.~Bellec}
\affiliation{Laboratoire de Physique de la Mati\`{e}re Condens\'{e}e, CNRS UMR 7336, Universit\'{e} de Nice Sophia-Antipolis, 06108 Nice, France}
\author{U.~Kuhl}
\affiliation{Laboratoire de Physique de la Mati\`{e}re Condens\'{e}e, CNRS UMR 7336, Universit\'{e} de Nice Sophia-Antipolis, 06108 Nice, France}\email{ulrich.kuhl@unice.fr}
\affiliation{Fachbereich Physik der Philipps-Universit\"{a}t Marburg, D-35032 Marburg, Germany}
\author{F.~Mortessagne}
\affiliation{Laboratoire de Physique de la Mati\`{e}re Condens\'{e}e, CNRS UMR 7336, Universit\'{e} de Nice Sophia-Antipolis, 06108 Nice, France}

\date{\today}

\begin{abstract}
Experiments on hexagonal graphene-like structures using microwave measuring techniques are presented. The lowest transverse-electric resonance of coupled dielectric disks sandwiched between two metallic plates establishes a tight-binding configuration. The nearest-neighbor coupling approximation is investigated in systems with few disks. Taking advantage of the high flexibility of the disks positions, consequences of the disorder introduced in the graphene lattice on the Dirac points are investigated. Using two different types of disks, a boron-nitride-like structure (a hexagonal lattice with a two-atom basis) is implemented, showing the appearance of a band gap.
\end{abstract}

%PACS 2008: 42.70.Qs Photonic bandgap materials (for photonic crystal lasers, see 42.55.Tv)
%PACS 2008: 73.22.-f Electronic structure of nanoscale materials: clusters, nanoparticles, nanotubes, and nanocrystals
%PACS 2008: 71.20.-b Electron density of states and band structure of crystalline solids
%PACS 2008: 42.25.Fx Diffraction and scattering
%PACS 2008: 03.65.Nk Scattering theory

%\keywords{Graphene, microwave experiments, Dirac point, edge states}

\maketitle

\section{INTRODUCTION}
\label{sec:Intro}
Graphene, due to its unusual and astonishing electronic and physical properties, is a subject of intense focus, especially since it was experimentally realized in 2005 by K. S.~Novoselov, A. K.~Geim and coworkers \cite{nov05}, who were honored by the Nobel prize in physics in 2010. One of graphene's amazing features is the linear dispersion relation close to the $K$-points and the Dirac like Hamiltonian whose consequences can be seen in the density of states (for details on graphene see the Revs.~\onlinecite{cas09,gei07}). Graphene is one of the candidates to take over from silicon-based electronics. Graphene has a macroscopic ballistic carrier transport due to its high charge carrier mobility and a high magnetic susceptibility $\mu$ \cite{sch07b}. A problem is still that graphene in its pure form is metallic and has no band gap $\Delta E$. Nevertheless, a band gap can be induced by using bilayer graphene \cite{zha09}, disorder \cite{per06,muc09}, lateral-superlattice, epitaxially grown graphene on top of a crystal-like boron nitride,\cite{for98,ber04,dea10} or by confinements \cite{ata10,hua10}. Applying stress to graphene can induce effects on the electron transport like a magnetic field \cite{lev10}. Thus graphene offers a big variety of interesting possibilities from both applied and fundamental points of view.

In real single layer graphene, it is quite difficult to prepare special shapes, induce precise structures of vacancies and so on. This can be easily performed in experiments with classical waves. For example in microwave analog experiments, the Dirac point \cite{zan10,bit10a,kuh10a}, a topological phase transition \cite{arXbel12}, the global band structures \cite{bit12}, and edge states \cite{kuh10a,bit12} have been observed. These states, which are located only at the zig-zag edges can lead to self-guiding unidirectional electromagnetic transport if the hexagonal lattice is created by ferritic scatterers \cite{poo11}. Edge states have also been found in acoustics \cite{zho11}.

In this paper, we mimic the established tight-binding form of graphene's Hamiltonian \cite{wal47,rei02,mat09} with a microwave setup using dielectric disks between two metallic plates. In a previous publication, we investigated the Dirac point in the density of states (DoS) \cite{kuh10a}. Here, we concentrate more on the experimental details of the realization of the tight-binding Hamiltonian (see Sec.~\ref{sec:Exp}). We show the flexibility of the experimental ansatz by investigating two different cases. The first case is to introduce positional disorder, where for strong disorder the Dirac point (see Sec.~\ref{sec:DisorderGraphene}) vanishes. The second case is to realize a boron-nitride-like structure by using two different kinds of disks on the lattice (see Sec.~\ref{sec:BoronNitride}). In this case, a band gap opens.

\section{THEORETICAL BACKGROUND}
\label{sec:Theory}
\begin{figure}
\includegraphics[width=.5\columnwidth]{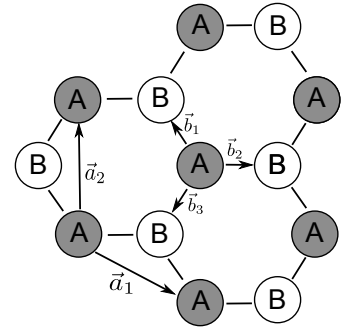}
\caption{\label{fig:graphene_lattice}
(Color online) Graphene as a superposition of two triangular sublattices A and B with lattice vectors $\vec a_1$, $\vec a_2$ and the vectors $\vec b_1$, $\vec b_2$, $\vec b_3$ connecting nearest neighbors.}
\end{figure}

Graphene is a two-dimensional crystal of carbon atoms ordered in a regular hexagonal lattice, a so-called ``honeycomb'' structure (see Fig.~\ref{fig:graphene_lattice}). It was first completely described theoretically by Wallace in 1947 \cite{wal47}. The far-reaching equivalence of graphene's Hamiltonian and the Dirac Hamiltonian for a free particle is depicted in Ref.~\onlinecite{Sem84}. Using the order and the notation of Ref.~\onlinecite{Sem84}, we first give a short repetition of a standard derivation of graphene's Hamiltonian to underline its equivalence to our experiment.

Graphene's characteristic honeycomb lattice structure is caused by the $sp^2$ bond of the carbon atoms resulting in three $\sigma$-bonds in the plane including an angle of $120^\circ$ and a free $p_z$ orbital perpendicular to the plane. The bonding length between the carbon atom is $a_b= 1.42$\,\AA. The basis vectors of the hexagonal lattice --- illustrated in Fig.~\ref{fig:graphene_lattice} --- are $\vec a_1 = \left(\sqrt{3}/2, -1/2\right)\,a$ and $\vec a_2 = \left(0, 1\right)\,a$, where $a=\sqrt{3}\,a_b$ is the lattice constant. The vectors connecting the two sublattices are $\vec b_1 = \left(1/2\sqrt{3}, 1/2\right)\,a$, $\vec b_2 = \left(-1/\sqrt{3}, 0\right)\,a$, and $\vec b_3 = \left(1/2\sqrt{3}, -1/2\right)\,a$. In the tight-binding approximation, we assume a basis of localized eigenfunction at the lattice sites imitating the $p_z$-orbitals. Thus we can write down the Hamiltonian $H$ in second quantization which includes the position energy and the hopping term.

For the position term $H_{\mathrm{pos}}$, we use a diatomic system (indicated in Fig.~\ref{fig:graphene_lattice} by A and B) placed on the two triangular lattices A and B, which is realized, e.\,g., in boron nitride. In the tight-binding model an atom of sort A only interacts with its three nearest neighbors that all belong to the lattice B (see Fig.~\ref{fig:graphene_lattice}). Second nearest-neighbor hopping is here neglected. The energy difference between electrons localized on lattice A and electrons on lattice B is called $\beta = (\epsilon_A - \epsilon_B)/2$. The associated part of the Hamiltonian can be written as
\begin{eqnarray} \nonumber
H_{\mathrm{pos}}
  &=& \sum \epsilon_A ~n(\vec A) + \epsilon_B ~n(\vec B) \\ \label{eqn:Hpos}
  &=& \beta \sum_{\vec A} [n(\vec A) - n(\vec A+ \vec b_1)]+ \underbrace{\frac{\epsilon_A+\epsilon_B}{2}}_{\equiv E_0}~N_{tot}
\end{eqnarray}
where $n(\vec A)$ and $n(\vec A+ \vec b_1)$ are the counting operators of electrons at positions $\vec A$ belonging to lattice A and at $\vec B = \vec A+ \vec b_1$ to lattice B, respectively. They can be written using the annihilation ($U$, $V$) and creation operators ($U^\dagger$, $V^\dagger$) as $n(\vec A) = U^\dagger(\vec A) U(\vec A)$ and $n(\vec A+ \vec b_1) = V^\dagger(\vec A+ \vec b_1) V(\vec A+ \vec b_1)$. $ \sum_{\vec A}$ is the sum over all atoms of lattice A. The additional term proportional to $E_0$ times the total number of electrons is only a constant and can be neglected in the further treatment. It corresponds to an energy shift to the mean energy of the two atoms (for graphene to the eigenenergy of one atom).

The second part of the Hamiltonian includes the hopping of electrons from one lattice site to one of the neighboring sites. The hopping parameter $\kappa$ describes the probability for this process and corresponds to the nearest-neighbor coupling, which is accessible in the measurements (see Sec.~\ref{subsec:2disk}):
\begin{equation} \label{eqn:Htrans}
H_{\mathrm{hop}}=\kappa \sum_{\vec A,i}~[U^\dagger(\vec A) V(\vec A + \vec b _i) +V^\dagger(\vec A+ \vec b _i) U(\vec A)]
\end{equation}
with $i =1,2,3$ for the three nearest neighbors in lattice B of an atom A. The spin of the electron is here neglected, as a spin is not existing in the microwave experiment. The entire Hamiltonian equals the sum of Eqs.~(\ref{eqn:Hpos}) and (\ref{eqn:Htrans}) and has the eigenenergies:
\begin{equation}\label{eqn:disprel_long}
E(\vec k) = \pm \sqrt{\beta^2 + \kappa ^2\left|e^{i \vec k \cdot \vec b_1}+e^{i \vec k \cdot \vec b_2}+e^{i \vec k \cdot \vec b_3}\right|^2}.
\end{equation}
For the case of graphene ($\beta = 0$), the two bands with plus and minus signs are touching each other at the roots of the sum of the three exponential functions. These are given at $\vec{K} = \frac{2\pi}{\sqrt{3}~a}(1,\sqrt{3})$ and $\vec{K}^\prime = -\vec{K}$, the corners of the Brillouin zone. The dispersion relation is approximately linear close to these $K$ points with the so-called Fermi velocity $v_f$ as the proportionality constant:
\begin{equation}\label{eqn:Disprel}
E(\vec K + \vec q) \approx \pm \underbrace{\frac{\sqrt{3}}{2}\,\kappa\,a}_{\equiv v_f}\,|\vec q| ~~~~~\mathrm{for}~|\vec q|\ll|\vec K|
\end{equation}
This equals the relativistic dispersion relation $E(\vec p) = \pm c |\vec p|$ for a massless particle, when $v_f$ plays the role of the speed of light, and leads to the common name ``Dirac points'' for $K$ and $K^\prime$.

The strong connection to the Dirac theory is, moreover, given by the equivalence of the Dirac Hamiltonian and a certain form of graphene's Hamiltonian, which can be derived via a basis transformation described in Ref.~\onlinecite{Sem84}.

From Eq.~(\ref{eqn:Disprel}), the density of states (DoS) in the vicinity of the Dirac points can be calculated:
\begin{equation} \label{eqn:DoS}
\rho(E) \approx \frac{4}{\sqrt{3}\pi}\frac{|E|}{\kappa^2},
\end{equation}
which is again linear.

In our experiments, the DoS (strictly speaking the {\it local} density of states, LDoS) is an accessible quantity. In our previous work, we have observed the linear dependence of the DoS close to the Dirac point \cite{kuh10a}. In this paper we want to investigate the stability of the Dirac points against disorder (see Sec.~\ref{sec:DisorderGraphene}).

If the on-site energies are different, i.e., $\beta \ne 0$ like in boron nitride, then a gap is generated at the Dirac point, where the gap width is given by $2\beta=\epsilon_A - \epsilon_B$. An analog to boron nitride is presented in Sec.~\ref{sec:BoronNitride}.

\section{EXPERIMENTAL REALISATION OF A TIGHT-BINDING SYSTEM}
\label{sec:Exp}

\subsection{Introducing microwave experiments}
Microwave experiments have been proven to be an attractive implementation to perform quantum analog measurements in the realm of ``quantum chaos'' for 20 years \cite{stoe90,stoe99}. Exploiting the formal analogy of the Helmholtz and the Schr\"{o}dinger equations, they have been used to investigate many different effects like eigenvalue distributions, nodal domains \cite{kuh07a,die08b}, affirmation of periodic orbit theory \cite{kol94a,lau06}, random matrix theory\cite{kuh08b,die08a}, and the random plane wave model \cite{kuh07b}. Quantum-mechanical currents and the visualization of chaotic wave functions in billiard systems \cite{sri91,ste92} have already been in the center of attention. Nowadays, the area of application became much wider and reaches beyond the first investigations on chaotic billiards. For instance, one-dimensional wave guides with introduced disorder have been suitable to check scattering theory phenomena. Effects of uncorrelated and correlated disorder have been observed \cite{kuh00a,kuh08a,die11a} and localization phenomena have been investigated in two-dimensional systems using dielectric disks \cite{lau07}. Photonic band gap materials (PBGM) (for a review, see, e.g.,~Ref.~\onlinecite{joa08}) assuming infinitely long cylinders have been suggested for two-dimensional tight-binding realizations \cite{lid98}. Effects of disorder have been investigated theoretically \cite{lid00b} and implemented successfully experimentally \cite{bay01}.

Transport of electrons in lattices is related to wave phenomena. As large interest in the special properties of the band structure of graphene occurred, microwave analog experiments have been performed on graphene-like structures using different kinds of realizations \cite{zan10,bit10a,kuh10a,bit12}. Here, we want to emphasize the possibilities of using dielectric disks with a high index of refraction \cite{kuh10a}. This new field offers a high flexibility to realize random or symmetric lattice structures and investigate its characteristics by observing the local density of states at different positions.

\subsection{Setup}

\begin{figure}
\includegraphics[width=.99\columnwidth]{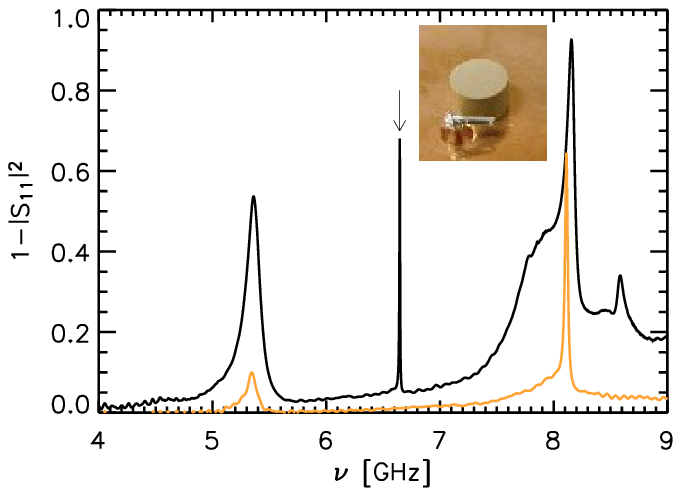}
\caption{\label{fig:SinglediskSpectrum}
(Color online) Spectrum of a single disk: black line corresponds to a measurement with the TE antenna shown in the inset; orange line shows the same measurement performed with a TM antenna. The arrow marks the first TE resonance that will be used for further experiments (from Ref.~\onlinecite{kuh10a}).}
\end{figure}

For the investigation of tight-binding systems we need wave functions that exists in a finite volume and decrease faster than $\frac{1}{|\vec r|^2}$ outside. This scenario can be considered as a potential well configuration. For the realization we chose dielectric disks (Temex-ceramics, E2000 series) with a dielectric permittivity $\epsilon_r \approx 36$ and a relative permeability $\mu_r = 1$. They have a diameter of 8\,mm and a height of 5\,mm (see inset of Fig.~\ref{fig:SinglediskSpectrum}).

These disks were previously used to realize disordered systems \cite{lau07}. They are placed on a flat copper plate. Atop another metallic plate is positioned parallel to the bottom plate. The two metallic plates are separated by a distance $h$. Thus the cutoff frequency for the first TE mode in air due to the plates is given by $\nu_{\mathrm{cut}}=c_0/(2h)$, where $c_0$ is the speed of light in air \cite{jac62}. The investigated resonance frequency of the disks is smaller than $\nu_{\mathrm{cut}}$ guaranteing that the TE waves are evanescent outside the disk. This assures additionally the two-dimensionality of the system \cite{stoe99}. Further descriptions about the disk eigenmodes are given in Sec.~\ref{subsec:disk_theory}.

In Fig.~\ref{fig:SinglediskSpectrum}, the reflection spectra are shown using two different kinds of antennas. The first is a pure dipole antenna, which consists of a vertical part only, exciting only TM modes (shown in orange). The second has an additional horizontal part that was soldered to the end of a standard dipole antenna (see inset) and can also excite the TE modes (spectrum shown in black). The sharp resonance marked by the arrow is the TE$_1$ resonance of the disk as it is not observable with a dipole antenna. Besides, it is nicely separated from the neighboring resonances and has a quite flat background. Thus it is a suitable resonance for further investigations. We always make sure that the other resonances are still separated from the frequency range of this TE resonance so that mixing or overlapping of TE and TM modes are avoided. Consequently, we work in a one-level regime, where each disk brings in only a single resonance.

\begin{figure}
\includegraphics[width=.99\columnwidth]{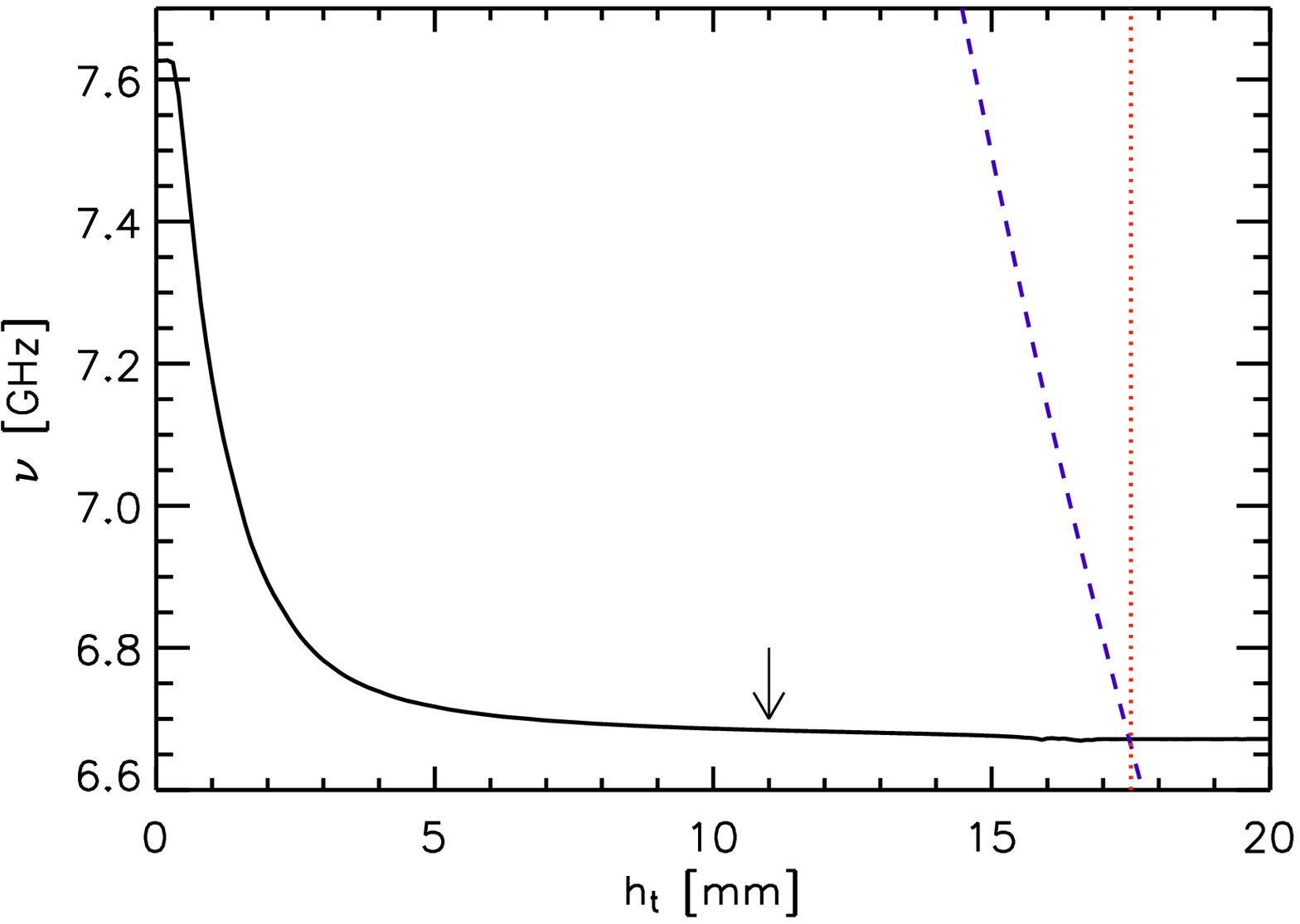}
\includegraphics[width=.99\columnwidth]{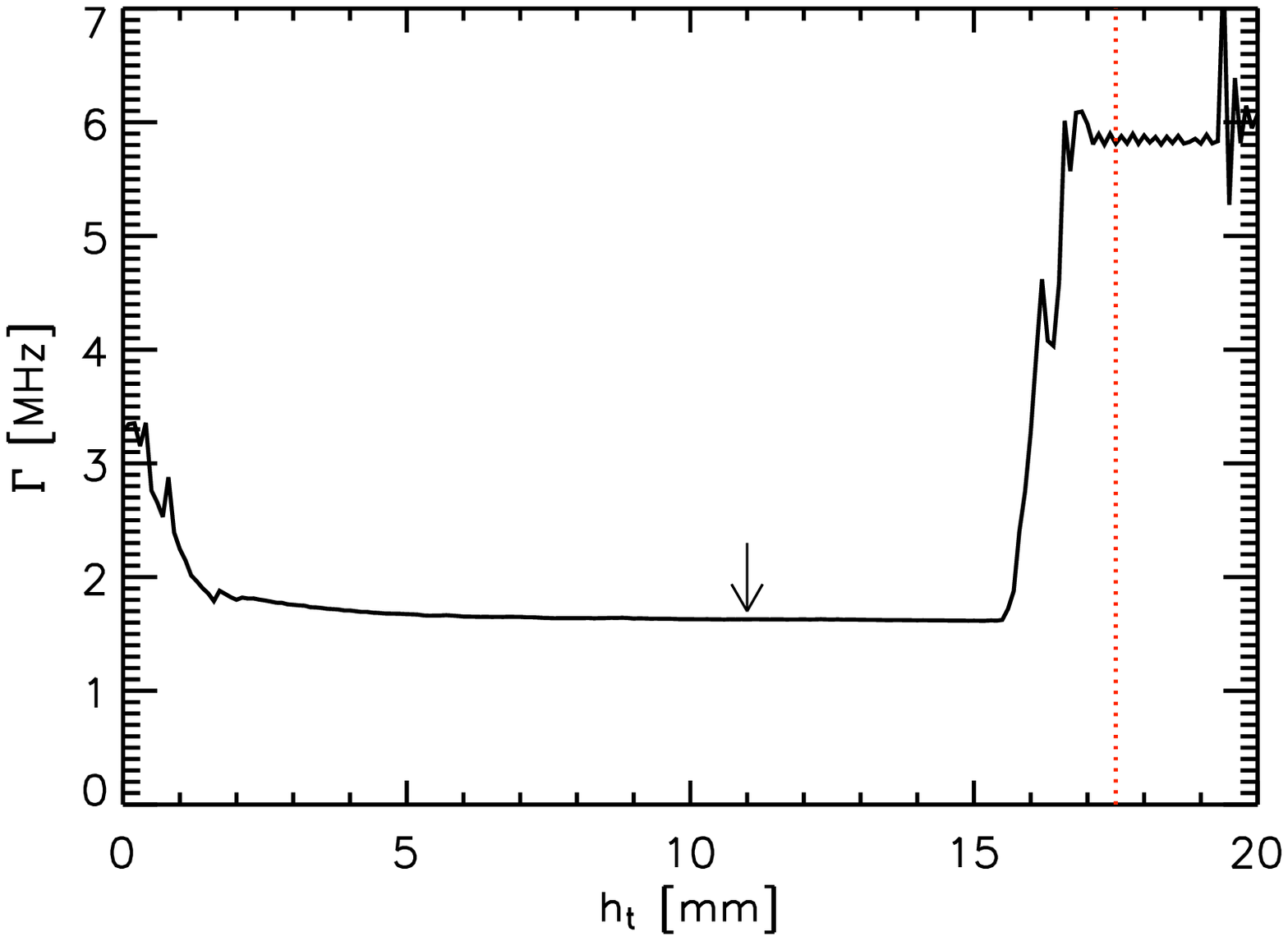}
\caption{\label{fig:ResonanceHeightDependence}
(Color online) Resonance frequency $\nu$ (upper) and width $\Gamma$ (lower) of a single disk are shown in dependence of the distance $h_t$ between the metallic top plates and the top of the disk. The dashed blue line corresponds to the cutoff frequency at the corresponding height and the vertical dotted red line indicates the height at which the resonance frequency is above the cutoff frequency. The arrow marks the chosen distance $h_t=11$\,mm between the top of the disk to the top plate for the experiments.}
\end{figure}

To mimic molecules or solids several disks have to be placed. As we want to mimic molecules consisting of one type of atoms, we have to ensure that each disk has the same eigenfrequency. One problem for this realization is the sensitivity of the resonance frequency once the top plate is very close to the top of the disks. This can be seen in Fig.~\ref{fig:ResonanceHeightDependence}, where the resonance frequency and width are shown in dependence of the height $h_t$. $h_t$ is the distance between the top of the disk to the metallic top plate. At $h_t$=0\,mm, the bottom plate is pressed against the upper plate using a step motor. Thereafter it is moved in steps of 0.1\,mm. At the beginning, this just corresponds to a release of the pressure and thus the resonance frequency does not change. Otherwise, for small heights the change of the resonance frequency is large for even tiny variations of the height and than saturates for larger heights. Assuming height variation along the setup of a few tenth of a millimeter and a typical resonance width of $\Gamma$= 2\,MHz, a height $h_t$ above 5\,mm ensures that the variation of resonance frequencies due to height variations stays within the resonance widths. Note that in Fig.~\ref{fig:ResonanceHeightDependence} the resonance frequency is given in GHz whereas the width is given in MHz. The width decays for small distances and stabilises also but increases strongly once the eigenfrequency gets closer to the cutoff frequency. For all further measurements, a height of $h$=16\,mm ($h_t$=11\,mm) is chosen to guarantee sharp and stable resonances.

The disks are positioned by a motorized $xy$-stage that is controlled by a PC programme. Hence it is possible to create structures of a few hundreds of disks with an accuracy of the order of 0.1\,mm. To ensure good and equivalent contact of the disks with the bottom plate the disks are pushed down while being placed.

\subsection{Disk eigenmodes}
\label{subsec:disk_theory}

The aim of our experiment is to realize a tight-binding system of localized functions with weak coupling to the outside. For the theoretical description, let us assume that $h_t$=0\,mm, i.\,e.\ the top plate touches the disk, the refractive index is $n=6$, and the disk height is $h=5$\,mm. The TE$_1$ mode consists only of the $z$ component of the magnetic field $\vec{B}=(0,0,B_z)$ and the perpendicular components of the electric field. For the given distance between top and bottom plates the cutoff frequency $\nu_{\mathrm{cut}} = 9.37$\,GHz for this first TE-mode. The $z$-dependence can be separated and if the resonance frequency is below the cutoff frequency the lowest TE resonance then looks like
\begin{equation}\label{eq:Bz}
B_z(x,y,z,k)=B_0 \times \left\{\begin{array}{rc}
\ \sin\left(\frac{\pi}{nh}z\right) J_0(k_\perp r); & r < r_D\\
\alpha \sin\left(\frac{\pi}{h}z\right) K_0(\gamma r); & r > r_D
\end{array}
\right.
\end{equation}
where $k_\perp=\sqrt{k^2-\left(\frac{\pi}{nh}\right)^2}$, $\gamma=\sqrt{\left(\frac{\pi}{h}\right)^2-k^2}$, and $r =\sqrt{x^2+y^2}$ is the distance from the center of the disk. $J_0$ and $K_0$ are Bessel functions and $\alpha$ is a constant to be determined from the continuity equations at the surface. At the resonance frequency of the disk, the wave number for $r>r_D$ is thus purely imaginary leading to an evanescent wave in the area outside the disk. Note that in the experiment there is a gap between the top plate and the disks, and this description is only valid approximately. We still use Eq.~(\ref{eq:Bz}) but with an effective $\gamma$ to be determined from the experiment.

\begin{figure}
\parbox{\columnwidth}{\includegraphics[width=.99\columnwidth]{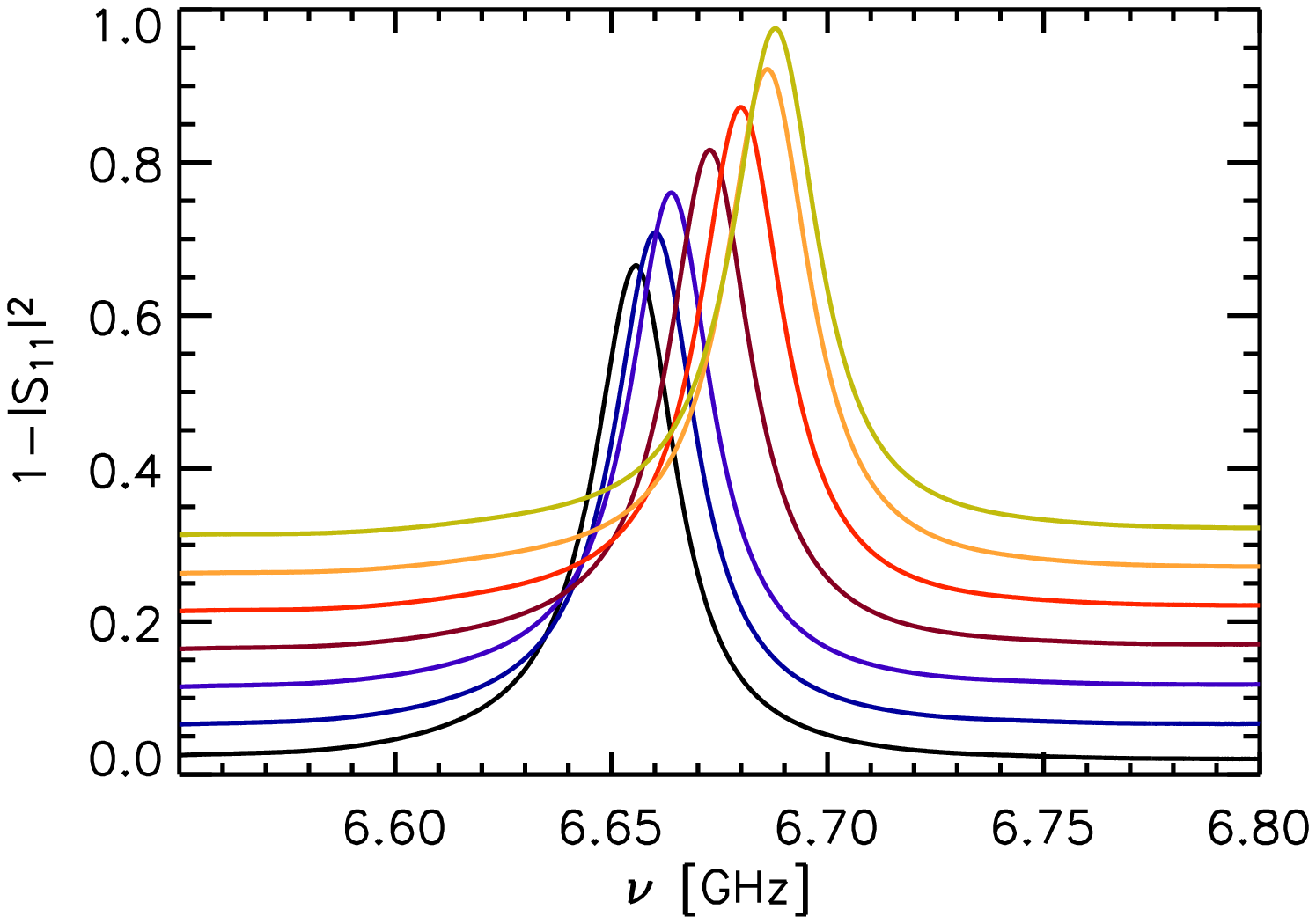}\\
\hspace*{.57\columnwidth}\raisebox{4cm}[0pt][0pt]{\includegraphics[width=.25\columnwidth]{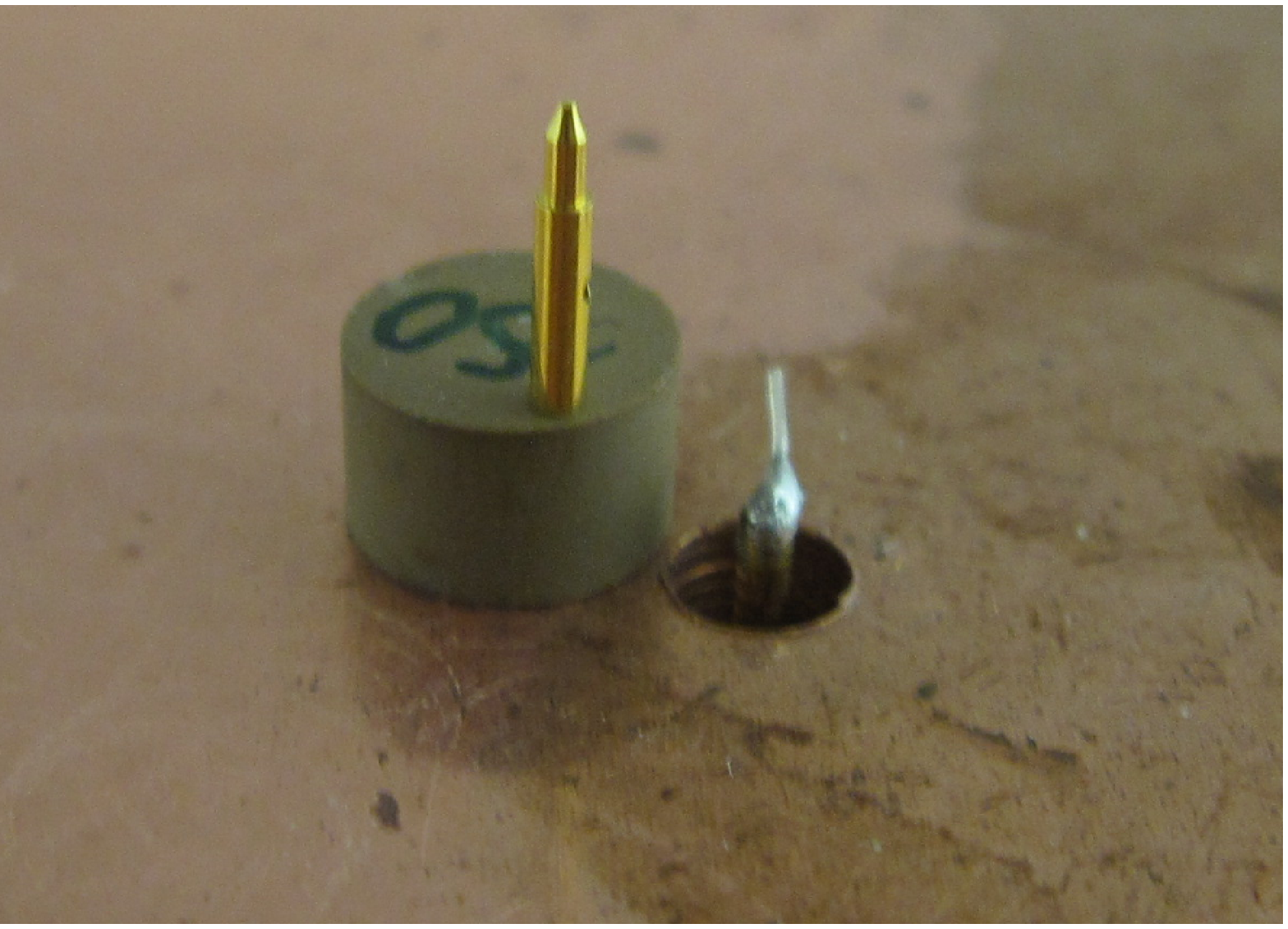}}\\}
\caption{\label{fig:Wavefct}
(Color online) The spectra of a single disk are shown for different perturber positions. All spectra are obtained without top plate. Lowest (black) spectrum: no perturber, next (blue) spectrum: perturber close to the border, the further up the spectrum is shifted, the closer is the perturber placed to the disk's center. The spectra are shifted in $y$ in steps of 0.05 for better visualization. The inset on the upper right shows a photograph of the disk including the metallic perturber and the measuring antenna.}
\end{figure}

To experimentally test the wave function inside the disk, we measured the reflection for a single disk as a function of the position of a perturber, i.e.\,a small metallic rod placed above the disk (see inset of Fig.~\ref{fig:Wavefct}). For this measurement the top plate is removed. The perturber induces a frequency shift, which is proportional to the intensity of the wave at the position of the perturber. In Fig.~\ref{fig:Wavefct}, a monotonous resonance shift is observed confirming that the wave function corresponds to the lowest Bessel function $J_0$. A direct coupling of the TE$_1$ mode to the TM$_0$ and TM$_1$ modes is strongly suppressed by the continuity conditions of the fields at the disk boundaries. Consequently, the wave function in the disks can be identified with an electron wave localized at a lattice site.

\subsection{Impact of the measuring antenna}
\label{subsec:antenna}

\begin{figure}
\includegraphics[width=.99\columnwidth]{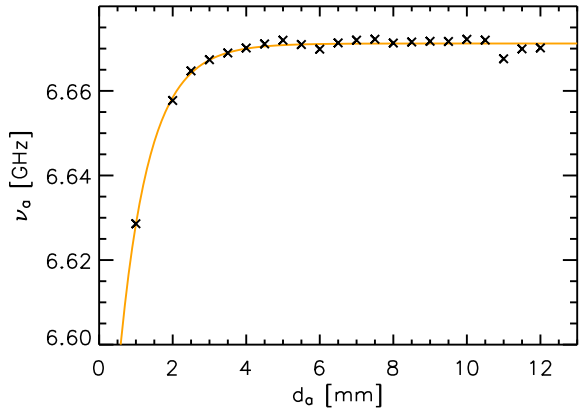}
\caption{\label{fig:sdisk_fit}
(Color online) Resonance positions $\nu_{\mathrm{exp}}$ versus distance $d_a$ between the disk and antenna (black crosses) with fit of Eq.~(\ref{distancetoantenne_fit}) (orange line) with $\kappa_a$=-13.89\,GHz, $\gamma$=0.52\,mm$^{-1}$, and $\nu_0$=6.67\,GHz.}
\end{figure}

For all the measurements we need to know the impact of the antenna on the disk eigenmode. Hence the coupling between two different objects, i.\,e. disk and antenna, is investigated. The pure presence of the antenna disturbs the symmetry of the system, but the coupling behavior can be quantified in a simple measurement series: we use a single disk with radius $r_D$ and increase its distance $d_a$ to the antenna. $d_a$ is the distance to the border of the disk and not to the center. Let us focus now on the shift of the resonance. The corresponding resonance frequencies are shown in Fig.~\ref{fig:sdisk_fit}. For smaller couplings, i.\,e.~for larger distances, the resonance frequency becomes higher and its amplitude decreases. The resonance frequency approaches the pure eigenfrequency for vanishing coupling, which thus equals 6.67 GHz. The shift of the resonance due to the antenna is proportional to the wave function's intensity at the position of the antenna, i.\,e.~at distance $r_D+d_a$. Thus we fit the function
\begin{equation}\label{distancetoantenne_fit}
\nu_a(d_a) = \kappa_a~|K_0\left[\gamma ~(r_D+d_a)\right]|^2+\nu_0
\end{equation}
to the experimentally measured resonance frequency $\nu_a$.

In Fig.~\ref{fig:sdisk_fit}, the experimental values are shown. A good agreement between fit (see orange line) and data points (crosses) is found. We obtained the fit parameters as $\kappa_a$=-13.89\,GHz, $\gamma$=0.52\,mm$^{-1}$, and $\nu_0$=6.67\,GHz. It's worth noting that $\kappa_a$ also includes the global coupling characteristics of the antenna. For an infinite distance the pure eigenfrequency $\nu_0 = 6.67$\,GHz remains and defines the eigenenergy of an isolated disk. Let us now examine the disk-disk coupling.

\subsection{Extracting the coupling parameter in two-disk measurements}
\label{subsec:2disk}

A system consisting of two disks with slightly different eigenfrequencies $\nu_0$ and $\nu_0+\Delta\nu_0$ and a distance depending coupling $\kappa_1(d)$ can be described by the Hamiltonian
\begin{equation}
H(d) = \left(\begin{array}{cc} \nu_0 & \kappa_1(d) \\ \kappa_1(d) & \nu_0+\Delta\nu_0 \end{array}\right)
\end{equation}
where the chosen basis $(1,0)$ and $(0,1)$ means ``electron'' in eigenstate of disk 1 with eigenfrequency $\nu_0$ and eigenstate of disk 2 with eigenfrequency $\nu_0+\Delta\nu_0$, respectively. The difference $\Delta\nu_0$ of the eigenfrequencies includes the influence of the antenna, quantified in Sec.~\ref{subsec:antenna}, and the discrepancy between the eigenfrequencies of the disks.

\begin{figure}
\includegraphics[width=.99\columnwidth]{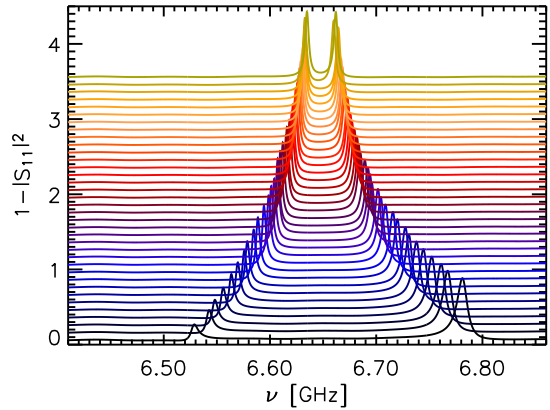}
\caption{\label{fig:TwodiskSpectra}
(Color online) Two-disk spectra for increasing distance. They are up-shifted with increasing distance. The lowest spectrum corresponds to a distance of 0.6\,mm, the topmost to 7.6\,mm.}
\end{figure}

\begin{figure}
\includegraphics[width=.99\columnwidth]{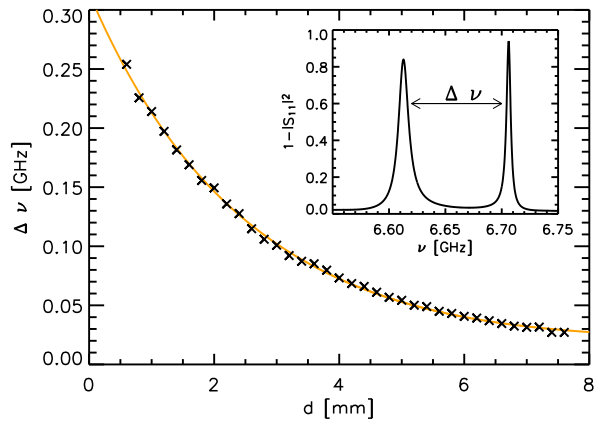}
\caption{\label{fig:TwoDoscDeltaNu}
(Color online) Resonance splitting of the two-disk system depending on the disk distance (black crosses). The experimental data agrees well with the fit function~(\ref{eq:deltanu}) for values $\kappa_0=1.35$\,GHz, $\gamma = 0.29$\,mm$^{-1}$, and $\Delta\nu_0=0.021$\,GHz. The inset shows exemplary the spectrum for $d=3.3$\,mm (from Ref.~\onlinecite{kuh10a}).}
\end{figure}

The Hamiltonian has the eigenvalues
\begin{displaymath}
\nu_{1,2}(d)=\nu_0(d)+\frac{1}{2}\Delta\nu_0 \pm \frac{1}{2}\sqrt{ 4~\kappa_1(d)^2+\Delta\nu_0^2}
\end{displaymath}
whose difference is given by
\begin{displaymath}
\Delta \nu(d) = \sqrt{4~\kappa_1(d)^2+\Delta\nu_0^2}
\end{displaymath}
The coupling parameter $\kappa_1$ can be estimated from the overlap of the wave function of the single disks. Using the evanescent waves [Eq.~(\ref{eq:Bz}) for $r>r_D$] outside the disks the coupling is approximately given by
\begin{equation}\label{eq:kappa1}
\kappa_1(d)=\kappa_0~\left|K_0~\left[\gamma(r_D+\frac{1}{2}~d)\right]\right|^2.
\end{equation}
For the shift of the resonance we thus obtain
\begin{equation}\label{eq:deltanu}
\Delta \nu(d) = \sqrt{4~\kappa_0^2\left|K_0\left[\gamma(r_D+\frac{1}{2}~d)\right]\right|^4+\Delta\nu_0^2}
\end{equation}
To check it experimentally we varied the coupling between two disks by increasing their distance in steps of 0.2\,mm and measured the splitting of the two resonances. Figure~\ref{fig:TwodiskSpectra} shows the corresponding spectra for distances of 0.6 to 7.6\,mm between the borders of the disks. With increasing distance, the decrease of the resonance splitting due to weaker coupling is clearly observable. After fitting the resonances with Lorentzians, we can extract the resonance splitting depending on the distances of the disks (see Fig.~\ref{fig:TwoDoscDeltaNu} crosses). The experimental findings are described by Eq.~(\ref{eq:deltanu}) and by fitting we obtained $\kappa_0$=1.35\,GHz, $\gamma$=0.29\,mm$^{-1}$, and $\Delta\nu_0$=0.021\,GHz.

In the next section we quantify the second nearest-neighbor coupling in a three-disk measurement.

\subsection{Extracting the next nearest-neighbor coupling parameter}
\label{subsec:3disk}

\begin{figure}
\includegraphics[width=0.99\columnwidth]{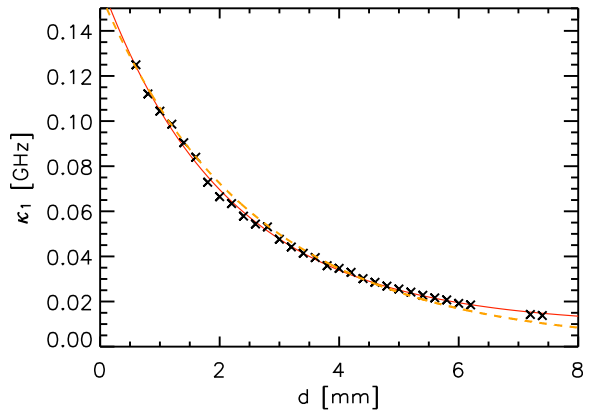}
\includegraphics[width=0.99\columnwidth]{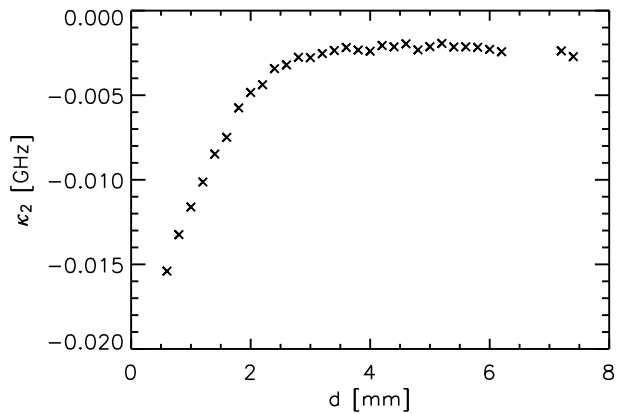}
\caption{\label{fig:ThreediskKappa12}
Coupling parameters for the three-disk measurement. Upper figure shows the nearest-neighbor coupling $\kappa_1$ with a fit of Eq.~(\ref{eq:kappa1}). The dashed line corresponds to the coupling parameter $\kappa_1$ from the two-disk measurement. The lower figure presents the extracted values for the next nearest-neighbor coupling $\kappa_2$.}
\end{figure}

\begin{figure}
\includegraphics[width=.99\columnwidth]{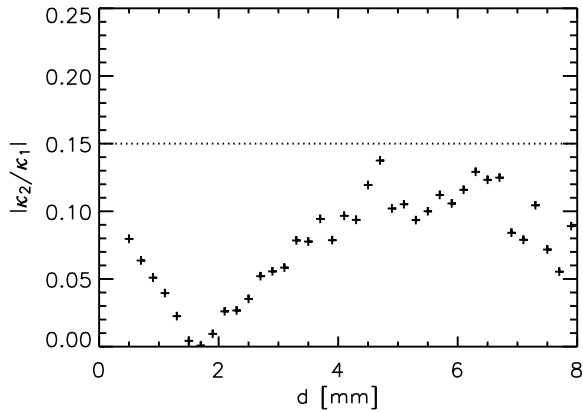}
\caption{\label{fig:3disk_kappas}
(Color online) Ratio of coupling parameters for increasing distance of the disk. The second-nearest-neighbor coupling reaches only less than 15\% of the nearest-neighbor coupling (horizontal line).}
\end{figure}

We performed an analogous measurement with three disks in a row whose distances will be equally increased in 0.2\,mm steps. The antenna remains close to the central disk. For simplicity, we neglect in the modeling of such a situation the influence of the antenna and assume perfectly equal disks. This leads to a Hamiltonian like
\begin{displaymath}
H(d) =  \left(\begin{array}{ccc} \nu_0&\kappa_1(d) &\kappa_2(d) \\ \kappa_1(d) & \nu_0 & \kappa_1(d) \\ \kappa_2(d) &  \kappa_1(d)&\nu_0 \end{array}\right)
\end{displaymath}
with $\kappa_1$ the nearest-neighbor coupling and $\kappa_2$ the second nearest-neighbor coupling. From its three eigenvalues $\nu_{1,2,3}$ the expressions for the coupling can be calculated:
\begin{eqnarray}  \nonumber
\kappa_1 (d) &=& \frac{\sqrt{2\nu_1^2+\nu_1 \nu_2- \nu_2^2-5 \nu_1 \nu_3+ \nu_2 \nu_3+2 \nu_3^2}}{3~\sqrt{2}}\\  \nonumber
\kappa_2 (d) &=& \frac{1}{3}~(\nu_1-2~\nu_2+\nu_3)
\end{eqnarray}
We measured the spectra of the equidistant three-disk system as a function of the border to border distance $d$ and extracted the corresponding eigenfrequencies. From those eigenfrequencies, we calculated the coupling parameters, which are presented in Fig.~\ref{fig:ThreediskKappa12}. Additionally, we plot the extracted coupling coefficient $\kappa_1$ from the two-disk measurement as dashed line. Overall the deviations are small. Note that we have neglected in the extraction of $\kappa_1$ for the three-disk system any deviations of the resonance frequencies $\Delta\nu_0$ of the disks. Fitting the next nearest-neighbor coupling with the same function as in Sec.~\ref{subsec:2disk}, we got
\begin{displaymath}
\frac{\kappa_1(d)}{\textrm{GHz}} =2.93~\left|K_0\left[0.36\left(4+ \frac{1}{2}\frac{d}{\textrm{mm}}\right)\right]\right|^2+0.009,
\end{displaymath}
which are similar to the fit values obtained for $\kappa_1$ in the two-disk experiment. The deviation of the fit values are coming from the different treatment of the differences of the eigenfrequencies of the disks $\Delta\nu_i$. In the two-disk problem, they are taken into account properly, whereas for the three-disk system, they are incorporated in the three fit constants. Note that in the limit for infinite distances, the coupling does not go to zero but to a constant value. The agreement of the coupling constants in case of the two- and three-disk systems underlines the applicability of the tight-binding approach. To justify the nearest-neighbor coupling approximation, we investigate the ratio $|\kappa_2/\kappa_1|$, which is presented in Fig.~\ref{fig:3disk_kappas}. The ratio is always smaller than 15\,\% thus the effects of next nearest-neighbor couplings are small but not negligible. While generally ignored in graphene, next-nearest-neighbor and even next-next-nearest-neighbor couplings are of the order of 5\% and can be even larger in bilayer or doped graphene \cite{rei02,mcc10,ben11}. Including higher-neighbor couplings can, for example shift the Dirac points or generate asymmetric band structures\cite{rei02,mcc10,ben11}.

In contrast to the PBGM, which are typically realized by infinitely long dielectric cylinders, the transport between the disks is evanescent, whereas in PBGM the waves are freely propagating between the dielectric cylinders\cite{lid98,lid00b,bay01}. Thus higher-order couplings are larger in the PBGM as the coupling is typically decreasing with $d^{-2}$, whereas in our case, it decreases with the modified Bessel function $|K_0(d)|^2$.

\subsection{Benzene -- the main component of graphene}
The next measurements were performed on a hexagonal cell. In the following, it will be called ``benzene'' though only the 6 $p_z$ orbitals of the carbon atoms are imitated and their hydrogen atoms are ignored. For symmetry considerations, it does not make any difference.

\begin{figure}
\includegraphics[width=.99\columnwidth]{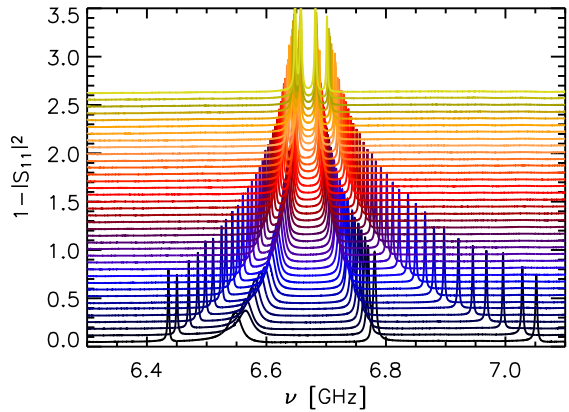}
\caption{\label{fig:benzene_spectrum} (Color online)
Spectra of hexagonal cells (benzene) for different distances. The spectra are up-shifted with increasing distance. The lowest spectrum corresponds to a distance of 0.5\,mm, increasing in steps of 0.2\,mm.}
\end{figure}

\begin{figure}
\includegraphics[width=.99\columnwidth]{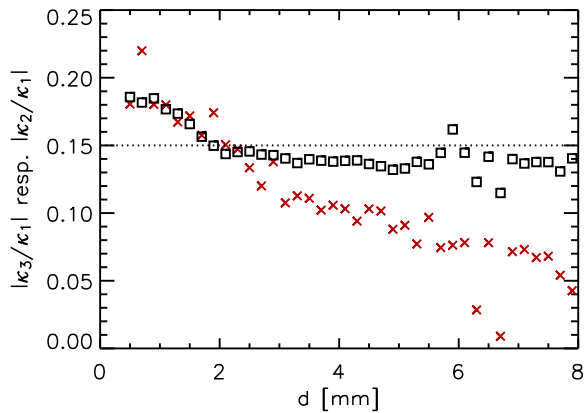}
\caption{\label{fig:benzene_kappas}
(Color online) Ratios $\kappa_3/\kappa_1$ (crosses) and $\kappa_2/\kappa_1$ (squares) vs the distance. From a distance of 2\,mm, the ratios are smaller than 0.15 (horizontal line).}
\end{figure}

First, we check the predictions of group theory: as benzene has the dihedral symmetry $D_{6h}$, we expect four resonances, consisting of two singlets and two doublets. The highest and lowest resonances are not degenerate. We measured spectra of benzene for different border to border distance $d$ of the neighboring disks increasing in steps of 0.2\,mm. The different spectra are shown in Fig.~\ref{fig:benzene_spectrum} and for each we find the expected four resonances. The degeneracy of the two central resonances was shown in Ref.~\onlinecite{kuh10a}. In case of benzene, each disk has two nearest neighbors ($\kappa_1$), two next-nearest neighbors ($\kappa_2$) and one next-next-nearest neighbors ($\kappa_3$). From the spectra, we extract the four resonances and calculated the coupling constants from the obtained resonances by
\begin{eqnarray}  \nonumber
\kappa_1 &=& \frac{1}{6}\left(-\nu_0-\nu_1+\nu_2+\nu_3\right),\\  \nonumber
\kappa_2 &=& \frac{1}{6}\left(+\nu_0-\nu_1-\nu_2+\nu_3\right),\\  \nonumber
\kappa_3 &=& \frac{1}{6}\left(+\nu_0+2\nu_1-2\nu_2+\nu_3\right).\\  \nonumber
\end{eqnarray}
We obtain a fit function of the following form for $\kappa_1$:
\begin{displaymath}
\frac{\kappa_1(d)}{\textup{GHz}} = 2.28~\left|K_0\left[0.33~\left(4+ \frac{1}{2}~\frac{d}{\textrm{mm}}\right)\right]\right|^2+0.0068.
\end{displaymath}
The $\kappa_1$ for the two-, three-disk and the benzene measurements shows only small deviations (typically below $\pm 0.01$\,GHz). Again, we have a look at the ratios of the three coupling parameters, which are illustrated in Fig.~\ref{fig:benzene_kappas}. Note that the ratios $\kappa_2/\kappa_1$ cannot be directly compared to the one obtained for the three-disk measurement as the next-nearest-neighbor distances are not the same. We observe that for distances larger than 2\,mm both ratios are smaller than 15\% (horizontal dashed line). As we want the influence of the higher-order neighbor couplings to be small, we consequently work with disk distances above 2\,mm and for the following measurements on graphene flakes, we use $d$=4\,mm.

Summarizing the basic measurements on few disks we conclude the following. The group theoretical predictions are fulfilled verifying the symmetry considerations. A consistent coupling behavior is found with small second and third nearest neighbor coupling. Thus we have realized a possibility to investigate tight-binding systems with a table top experiment, where several parameters can be varied easily. Now we proceed with hexagonal lattices.

Experimental results on graphene using the same setup have already been published in Ref.~\onlinecite{kuh10a}, where the Dirac point including the linear dispersion relation close to it have been observed. Additionally, edge states at zigzag edges and the lack of edge states at armchair edges have been shown. Here, we will proceed by showing the flexibility of the experiment and investigate the effect of positional (coupling) disorder on the Dirac point. Additionally, it is possible to use different kinds of disks, where we will realize a correspondence to boron nitride by using two types of disks on the two triangular sublattices.

\section{LOCAL DENSITY OF STATES}

As we want to relate our findings for larger systems to the density of states, we derive here the relation between the measured reflection signal $S_{11}$ at an antenna to the LDoS.

In quantum mechanics, the LDoS $\mathcal{L}(\vec r,E)$ is defined by
\begin{equation}
\mathcal{L}(\vec r,E) \equiv \sum _n |\phi_n(\vec r)|^2~\delta(E-E_n)
\end{equation}
where $\phi_n(\vec r)$ is the $n^{th}$ eigenfunction at position $\vec r$ and can be expressed via the Green function $G(\vec r,\vec r, E)$ of the system:
\begin{eqnarray}\nonumber
\mathcal{L}(\vec r,E)
  &=& -\frac{1}{\pi}~\textrm{Im}\left( \lim_{\epsilon \rightarrow 0} \sum_n \frac{|\phi_n(\vec r)|^2}{E-E_n+i\epsilon}\right) \\
\label{eqn:LDoS_G}
  &=& -\frac{1}{\pi}~\textrm{Im}\left[ G(\vec r,\vec  r,E) \right].
\end{eqnarray}

To relate the Green function to the reflection $R^2$, let us start with the probability $T^2$ for the incident wave to be transmitted via the antenna into the system and escape the system by other means, i.e. not via this antenna. In our case this corresponds to the absorption mainly in the top and bottom plate but also within the disks. By energy conservation, it is connected to the reflection $R^2$ and given by
\begin{eqnarray} \nonumber
1   &=& R^2+T^2 = |S_{11}|^2 + T^2, \\ \label{eqn:reflection_current}
T^2 &=& 1-|S_{11}|^2 = \oint \vec j ~d\vec s,
\end{eqnarray}
where $d\vec s$ describes a contour around the antenna (assuming a two-dimensional setup). $\vec j$ is the two-dimensional quantum-mechanical probability current,
\begin{displaymath}
\vec j = \frac{\hbar}{m} \textrm{Im}~(\psi^* \nabla_{xy}\psi)
\end{displaymath}
which in electrodynamics corresponds to the two-dimensional Poynting vector. For TE modes it is given by\cite{jac62}
\begin{equation} \label{eqn:poynting}
\vec S_{xy} =-\frac{ \mu^2}{\epsilon ~\omega }~\textrm{Im}~(B_z^* \nabla_{xy} B_z).
\end{equation}
We neglect the prefactors because only the proportionality of the current to $\textrm{Im}(\psi^* \nabla_{xy}\psi)$ is important. Assuming a point-like source at $\vec R$, the wave function $\psi = B_z$ is proportional to the Green function $G(\vec r,\vec R,E)$, as both are the solution of the Helmholtz differential equation with delta inhomogeneity. The proportionality coefficient $\Lambda(E,\vec R)$ depends on the coupling of the antenna to the system. Combining Eqs.~(\ref{eqn:LDoS_G})-Eq.~(\ref{eqn:poynting}), we obtain:
\begin{eqnarray} \nonumber
1-|S_{11}|^2
  &\propto& -|\Lambda(E,\vec R)|^2\oint \\ \nonumber
  &       & \times \textrm{Im}\left[G^*(\vec r,\vec R,E) \nabla_{xy} G(\vec r,\vec R,E)\right]\,\textrm{d}\vec s \\ \nonumber
  &\propto& |\Lambda(E,\vec R)|^2~\textrm{Im}~\xi_\beta(\vec R,E)  \\ \nonumber
  &\propto& |\Lambda(E,\vec R)|^2~\mathcal{L}(\vec R,E),
\end{eqnarray}
where the renormalized Green function
\begin{equation} \label{renorm Gf}
\xi_\beta(\vec R,E) =\lim_{\vec r\rightarrow \vec R} \left[G(\vec r,\vec R,E)-\frac{1}{2\pi}\ln\left(\frac{|\vec r-\vec R|}{\beta}\right) \right]
\end{equation}
removes the singularities at the antenna positions $\vec R$. $\beta$ is the scattering length of the antenna. A detailed derivation of the renormalized Green function can be found in Ref.~\onlinecite{tud08}.

It is not clear how the coupling parameter $\Lambda(E,\vec R)$ depends on $E$ or even on the LDoS. Therefore it is necessary to find a proper coupling theory. With the methods developed in Ref.~\onlinecite{tud08}, it is possible to derive an approximative description of the coupling for small cylindrical antennas showing an almost frequency independent prefactor $\Lambda(E,\vec R)$. Hence we finally obtain the relation between the LDoS and the reflection that is given by
\begin{equation}\label{eq:LDosS11}
\mathcal{L}(R,E) \propto 1-|S_{11}(R,E)|^2
\end{equation}
Based on this knowledge we will plot in the following figures $1-|S_{11}|^2$ on the $y$-axis.

\section{DISORDERED GRAPHENE}
\label{sec:DisorderGraphene}

\begin{figure}
\includegraphics[width=.99\columnwidth]{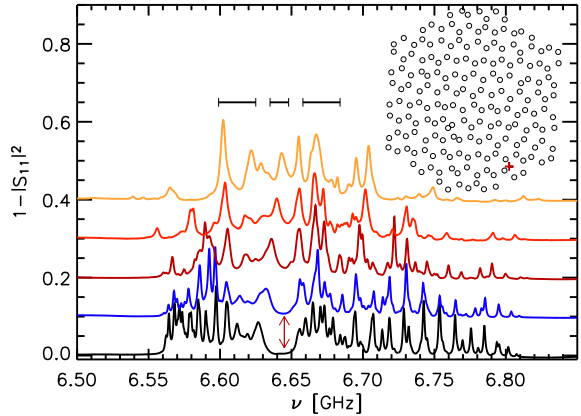}
\caption{\label{fig:graphene_noise}
(Color online) The inset on the upper right shows the graphene flake of hexagonal shape with maximal deposition of $\sigma$=4\,mm. The antenna position is marked by the red cross.
The spectra of the hexagonal graphene flakes with different maximal random depositioning $\sigma$ are shown. The lowest (black) spectrum corresponds to the unperturbed one ($\sigma$=0\,mm). The spectra with $\sigma$=1, 2, 3 and 4\,mm are shifted upwards correspondingly.  The bars indicate the region of averaging used in Fig.~\ref{fig:LDoS_noise}.}
\end{figure}

Graphene's astonishing DoS is a consequence of its highly symmetric hexagonal lattice structure. But in the real world the perfect hexagonal symmetry of graphene will be distorted by different means. For example, external strain \cite{lev10}, vacancies \cite{per06}, matching to substrates \cite{li10}, edges, and corners will destroy the symmetry. Additionally, they can introduce disorder, e.g., via the position of vacancies in the bulk \cite{per06} or via the edges in finite systems \cite{muc09}. Effects of Anderson localization have been numerically observed in transport through graphene nanoribbons with rough surfaces \cite{lib12}. In a work on PBGM, it was shown that a band gap can be quite stable against positional disorder if the original band structure comes from the Mie resonances \cite{lid00b}. This corresponds to our experimental realization apart from the fact that we have two touching bands at the Dirac point and not a proper gap.

One kind of perturbation that can be easily modeled by our setup is bulk disorder within the disk position. Starting from the perfect graphene lattice position of the disks $\vec{r}_{i,0}$, we create a new position $\vec{r}_{i,\sigma}=\vec{r}_{i,0} + \Delta r_i (\cos\theta_i e_x + \sin\theta_i e_y)$, where  $\Theta_i$ is randomly chosen from \{0,2$\pi$\} giving the direction of the shift and $\Delta r_i$ is randomly chosen from \{0,$\sigma$\} giving the length of the shift. $e_x$ and $e_y$ are the unit vectors in $x$ and $y$ directions. The positional disorder corresponds in the tight-binding description to a disorder in the coupling.

We want to test how stable the Dirac point is against an introduced disorder. To reduce effects of edge and corner states we chose a hexagonal graphene flake as shown in the inset of Fig.~\ref{fig:graphene_noise}. The antenna position where the reflection $S_{11}$ is measured is marked by a red cross. Starting with a regular lattice with a disk to disk distance of $d$=4\,mm, we add noise to every disk position as described above. We realized four disordered samples with a maximal distance to the original position of $\sigma=$\, 1, 2, 3, and 4\,mm. The inset of Fig.~\ref{fig:graphene_noise}, corresponds to the maximal value of $\sigma$=4\,mm. The behavior of the LDoS is shown in Fig.~\ref{fig:graphene_noise}. For $\sigma=$\,1\,mm we find a quite similar spectrum and can still see a strong reminiscence of the Dirac point in the LDoS (red arrow). For bigger shifts the structure disappears and the Dirac point becomes unobservable.

\begin{figure}
\includegraphics[width=.99\columnwidth]{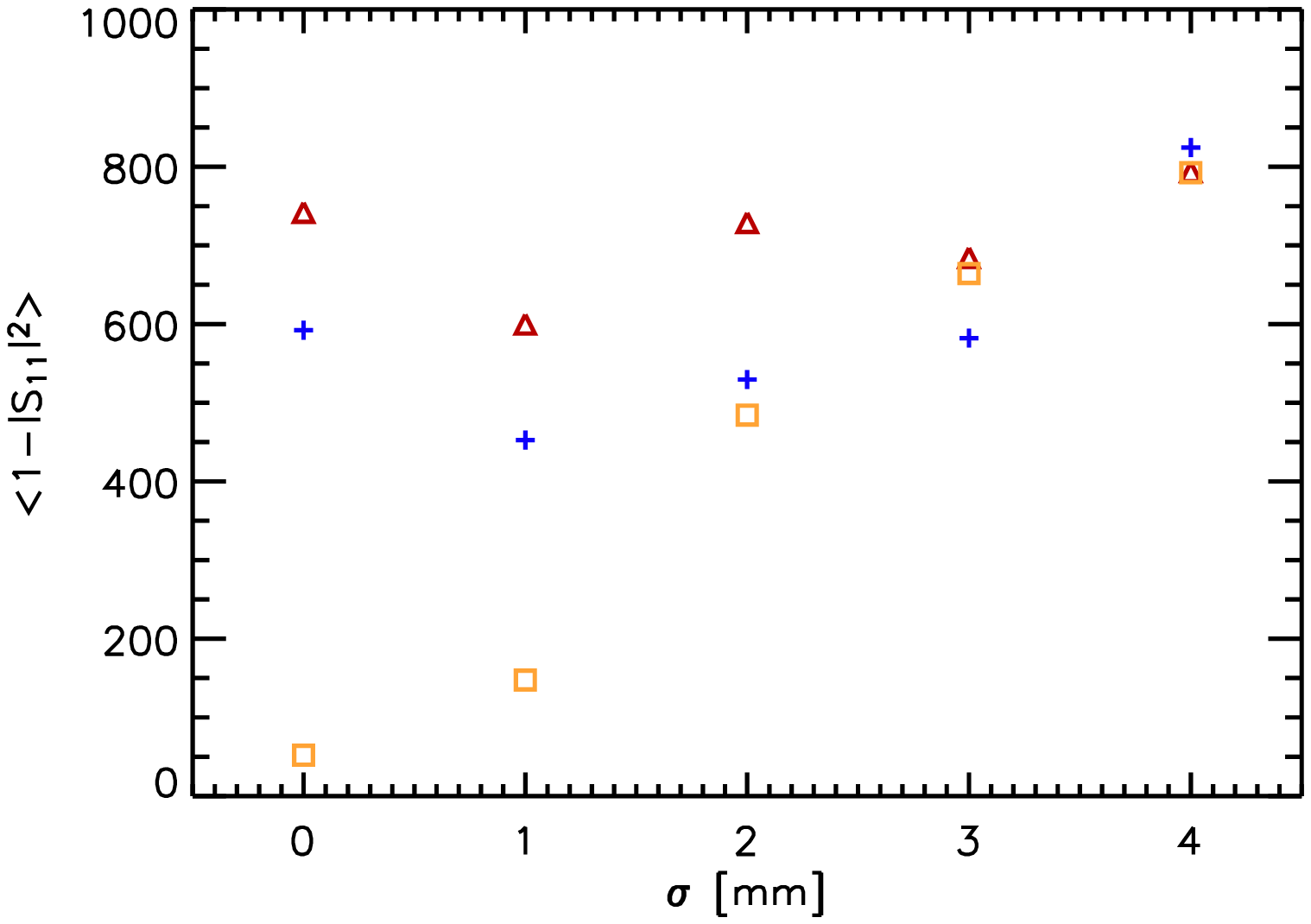}
\includegraphics[width=.99\columnwidth]{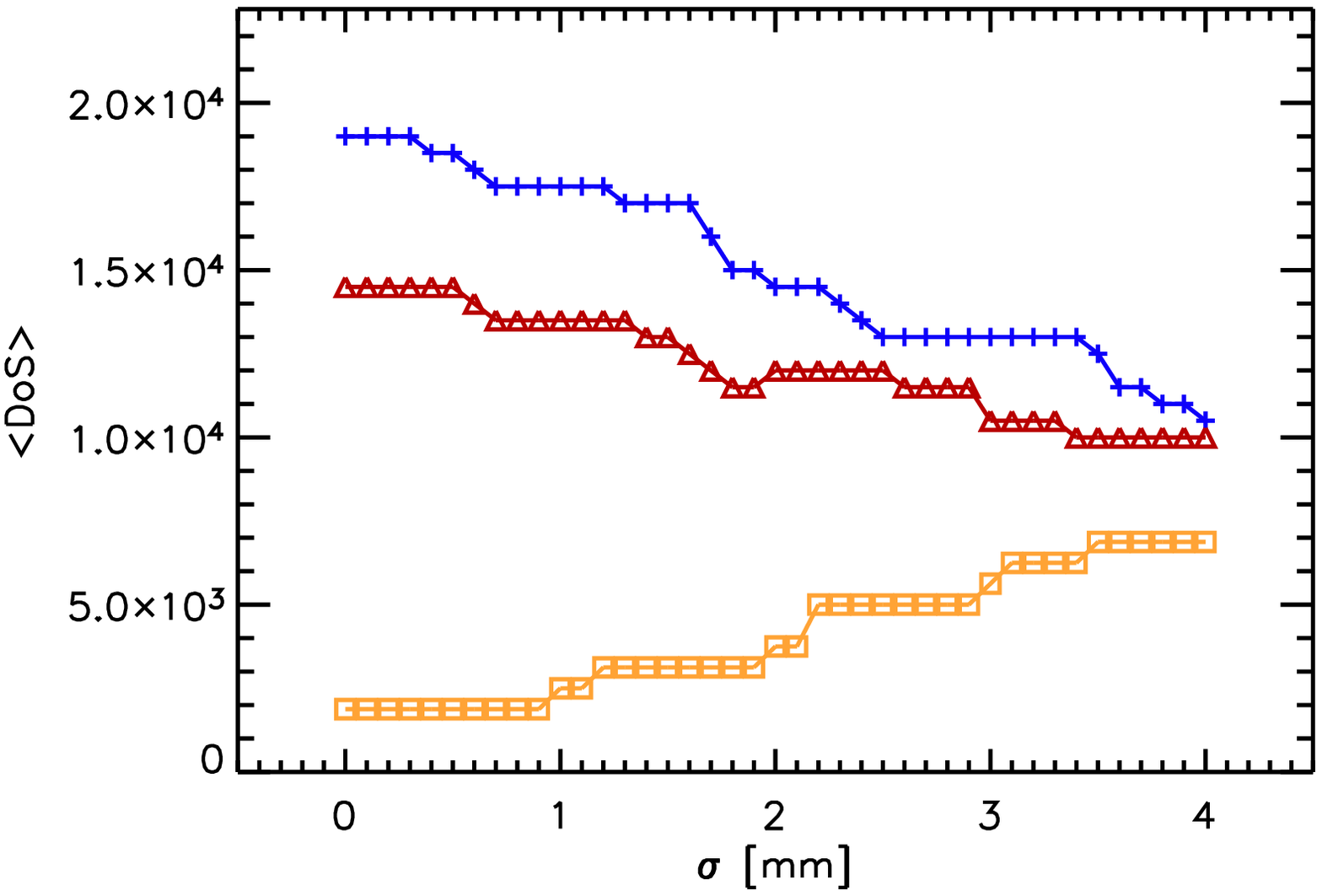}
\caption{\label{fig:LDoS_noise}
(Color online) On the upper figure the frequency averaged local density of states $\langle 1-|S_{11}|^2\rangle$ around the Dirac point (squares) and within the bands at lower (crosses) and higher (triangles) frequencies than the Dirac point are shown for different maximal disorder $\sigma$. The windows for averaging are indicated in Fig.~\ref{fig:graphene_noise}.
On the lower part the numerically obtained averaged DoS is presented. Symbols are used correspondingly to the upper figure.}
\end{figure}

To investigate this in a more quantitative manner, we perform averages over small frequency windows. The averaging windows are indicated by bars in Fig.~\ref{fig:graphene_noise} and are 13\,MHz at the Dirac point and 26\,MHz inside the bands. To observe effects of the Dirac points in the averaged quantity, the averaging range around the Dirac point is chosen so small that a reduction of the DoS is still clearly observed. It is necessary to introduce the average as we are dealing with finite systems. Thus the experimental value at the Dirac point is dominated by the frequency distance of the closest resonance and its width. If small disorder is introduced, we are not testing the stability of the global structure of the Dirac point but only the stability of the closest eigenvalue. By averaging locally, we wash out this effect but still keeping the information of the diminution of the DoS close to the Dirac point.

In Fig.~\ref{fig:LDoS_noise}, on the upper part the averaged LDoS ($\langle 1-|S_{11}|^2\rangle$) is presented around the Dirac point (squares) and within the two bands (triangles and crosses). One observes with increasing disorder an increase at the Dirac point, whereas the averaged LDoS within the bands oscillates. At $\sigma$=2\,mm, the averaged values are of the same order, thus indicating the final vanishing of the Dirac point. That the Dirac point is quite stable against disorder perturbations is in agreement with the findings for PBGM \cite{lid00b}.

Additionally, we performed numerical tight-binding simulations on the same flake as in the experiment including more than only nearest-neighbor couplings. We diagonalized the Hamiltonian
\begin{equation}\label{tb-matrix}
H_{\mathrm{tb}}= \nu_0~\mathds{1} + \mathbb{K}
\end{equation}
where $\nu_0=6.65$\,GHz is the eigenfrequency of a single disk. The coupling matrix $\mathbb{K}$ contains the coupling elements $\kappa$. A disk is coupled to all neighbors that have a border to border distance $d$ less than 18\,mm. The coupling parameter $\kappa$ is calculated by Eq.~(\ref{eq:kappa1}) using the specific disk distances $d$. The eigenfrequencies are obtained by diagonalizing the Hamiltonian. Then the DoS is calculated. On the lower part in Fig.~\ref{fig:LDoS_noise} the averaged DoS for different frequency ranges are shown. The average is performed over the same frequency windows as in the experimental case. In contrast to the experiment, the DoS is shown and not the LDoS. One observes also here that the density close to the Dirac point is increasing. Each step actually corresponds to one additional state inside the averaging region. Within the bands, a decrease of the averaged density of states is found. Overall, the Dirac point seems to be quite stable against the introduction of small white noise disorder.

\section{IMITATING BORON NITRIDE}
\label{sec:BoronNitride}

Boron nitride has a hexagonal lattice structure, but with two different sorts of atoms placed on the two triangular sublattices. Double layer of boron-nitride graphene lattices are a candidate to have on the one hand a band gap of similar energy than silicon and on the other hand keep the high charge carrier mobility of graphene \cite{han07}. Thus it is one candidate to substitute silicon and overcome especially problems creating high frequency transistors.

\begin{figure}
\includegraphics[width=.99\columnwidth]{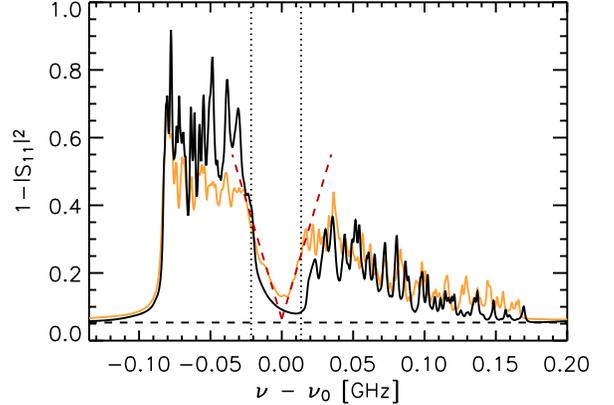}
\caption{\label{fig:boron-nitride}
(Color online) Two spectra of the same lattice configuration, but consisting of one type of disks (graphene, orange), respectively, two types of disks (boron-nitride, black). The disk close to the measuring antenna was not moved, to guarantee a consistent coupling. The dotted lines correspond to the eigenfrequencies of the two types of disks and indicate the band gap size. The spectra were shifted by its corresponding Dirac frequency $\nu_0$ such that the center of the bands are the same. Additionally, the red dashed lines indicate the linear behavior of the DoS (see also Ref.~\onlinecite{kuh10a}). The dashed horizontal line corresponds to the background signal of the experiment.}
\end{figure}

The energy difference $\beta$ of the two types of atoms causes a widening of the band gap between the conducting and the valence band, corresponding to Eq.~(\ref{eqn:disprel_long}). We imitate this situation using two charges of disks with mean eigenfrequency of $\nu_A$=6.637\,GHz and $\nu_B$=6.672\,GHz and create the same lattice once with equal disks and once with the different disks. A comparison of the resulting LDoS is presented in Fig.~\ref{fig:boron-nitride}. The orange curve, already presented in Ref.~\onlinecite{kuh10a} [see Fig 4(a)], corresponds to a graphene lattice with the Dirac point at $\nu_A$=6.637\,GHz. The curves were shifted by their corresponding Dirac frequencies $(\nu_0)$ so that the central frequency for graphene and boron nitride coincides. The dashed horizontal line corresponds to the experimental background signal, mainly coming from the coupling of the antenna to the TM$_0$ mode and thus corresponds to the 0 for the DoS. The red dashed lines indicate the linear behavior of the DoS close to the Dirac point. The black curve shows the LDoS in case of boron-nitride. A gap and a corresponding shift of the gap are observed. The shift is due to the fact that we use here two different disks. Thus the center of the band is expected to be at $\frac{1}{2}~(6.637+6.672)\,\mathrm{GHz}=6.655\,\mathrm{GHz}$, which is in good agreement with the experiment. The gap width is expected to be $\Delta\nu_g=|(6.637-6.672)|\,\mathrm{GHz}=0.035\,\mathrm{GHz}$, which corresponds to the difference between the dotted lines. It is in agreement with the experimentally observed gap.

\section{CONCLUSIONS}
\label{sec:conclusions}

In this paper, we present experimental results of hexagonal tight-binding configurations in a microwave setup. First, we show in detail how a tight-binding setup is realized. For setups with few disks, we compare next-nearest-neighbor coupling to the nearest-neighbor coupling. The flexibility of the experiment allows to vary easily disks positions, i.e.\,the coupling, but also the ``on-site energies,´´ i.e.\,eigenfrequency of the disks. We choose exemplarily to investigate disordered graphene. The disorder measurements show that the Dirac points seems to be quite stable against white noise disorder. As a second example showing the variation of the on-site energy, we observe a band gap by using two different type of disks corresponding to a boron nitride sheet.

Exploiting the flexibility of the setup the realization of many different tight-binding structures are imaginable. Realizing different kinds of vacancies including disorder, transport properties along lattice axes. By lattice deformation, one can either play with $K$ and $K'$ points \cite{mon09a,mon09b}, which have been experimentally realized using a similar setup \cite{arXbel12}, or simulate pseudo-magnetic fields to generate associated Landau levels \cite{gui10b}.

\begin{acknowledgments}
This work was supported by the Deutsche Forschungsgemeinschaft via an individual grant and the Forschergruppe 760: Scattering systems with complex dynamics. S.~B.\ thanks the LPMC at Nice for the hospitality during several long term visits and the University of Nice and the F\'{e}d\'{e}ration D\"{o}blin for financial supports.
\end{acknowledgments}

\end{document}